\documentclass[journal]{IEEEtran}

\usepackage[noadjust]{cite}
\usepackage[colorlinks=true,citecolor=green,linkcolor=red,urlcolor=blue]{hyperref}
\usepackage{times}
\usepackage{epsfig}
\usepackage{graphicx}
\usepackage{amsmath}
\usepackage{amssymb}
\usepackage{verbatim}
\usepackage{booktabs} 
\usepackage{adjustbox}
\usepackage{footnote}
\usepackage{colortbl} 
\usepackage{xcolor} 
\usepackage{xfrac}
\usepackage[ruled,vlined,noresetcount]{algorithm2e}

\usepackage[utf8]{inputenc}
\usepackage[english]{babel}
 
\usepackage{amsthm}

\begin{document}
\date{\today}

\title{On the Construction of Polar Codes for Achieving the Capacity of Marginal Channels}

\author{Amirsina~Torfi$^{*}$, Sobhan~Soleymani$^{*}$,~\IEEEmembership{Student Member,~IEEE}, Siamak~Aram$^{\star}$, and Vahid~Tabataba Vakili$^{\dagger}$,~\IEEEmembership{Member,~IEEE}\\
$^{*}$Department of Computer Science and Electrical Engineering, West Virginia University\\
$^{\star}$Department of Epidemiology and Public Health, University of Maryland Baltimore County\\
$^{\dagger}$School of Electrical Engineering, Iran University of Science and Technology}

\maketitle


\begin{abstract}
Achieving security against adversaries with unlimited computational power is of great interest in a communication scenario. Since polar codes are capacity achieving codes with low encoding-decoding complexity and they can approach perfect secrecy rates for binary-input degraded wiretap channels in symmetric settings, they are investigated extensively in the literature recently. In this paper, a polar coding scheme to achieve secrecy capacity in non-symmetric binary input channels is proposed. The proposed scheme satisfies security and reliability conditions. The wiretap channel is assumed to be stochastically degraded with respect to the legitimate channel and message distribution is uniform. The information set is sent over channels that are good for Bob and bad for Eve. Random bits are sent over channels that are good for both Bob and Eve. A frozen vector is chosen randomly and is sent over channels bad for both. We prove that there exists a frozen vector for which the coding scheme satisfies reliability and security conditions and approaches the secrecy capacity. We further empirically show that in the proposed scheme for non-symmetric binary-input discrete memoryless channels, the equivocation rate achieves its upper bound in the whole capacity-equivocation region.
\end{abstract}
\section{Introduction}
A cryptosystem is information-theoretically secure if it has no information leakage. It means that Eve with unlimited computational power could not break the system. An encryption protocol in which information-theoretic security is considered is not vulnerable to developments in computational power.
 
An example of an information-theoretically secure system is the one-time pad. Information-theoretically secure communication was introduced in 1949 by Shannon. He proved that one-time pad cryptosystem is secure~\cite{sha49}.
 
An encryption algorithm is perfectly secure if its output cipher-text provides no information about the plain-text when the key is not known. If $E$ is a perfectly secure encryption function, for any fixed message \textit{M} there must exist at least one key for each cipher-text \textit{c}, such that $c=E_{k}(M)$. It has been proven that any encryption algorithm with perfect secrecy must use keys similar to one-time pad keys~\cite{sha49}.

There are a wide variety of cryptographic tasks that implement information-theoretic security as a useful and meaningful concept. Some of them are mentioned below:
\begin{itemize}
\item Shamir's secret sharing algorithm is information-theoretically secure, which splits the secret into pieces and gives each piece to a specific person. To regenerate the secret at least a portion of pieces is required, otherwise, reconstruction of the secret is impossible.
\item Quantum cryptography is a very important part of information-theoretic cryptography.
\end{itemize}

A weaker notion of security named physical layer encryption defined by Wyner established a large area of research. It uses the physical channel for its security by signal processing and coding techniques. This notion of security is provable, unbreakable, and quantifiable.

Wyner's initial physical layer encryption work assumes that Alice wants to send a secret message to Bob without Eve being able to decode it. It was shown that if the channel from Alice to Bob is statistically better than the channel from Alice to Eve, secure communication is possible~\cite{wyn75}. As Alice tries to transmit messages to Bob through a communication channel, her transmissions may reach the adversary Eve through another the wiretap channel.

Encoder maps \textit{k-bit} message \textit{M} to codeword \textit{X} and sends it on the channel. Bob receives \textit{Y} on the main channel $W_m$, while on the wiretap channel $W_w$ Eve receives \textit{Z}. Decoder maps \textit{Y} to $\hat M $. A reliable and secure system should be achieved when message length tends to infinity. 
\begin{eqnarray}
\textbf{Reliability}: \mathop {\lim }\limits_{k \to \infty } \Pr (M \ne \hat M) = 0\
\end{eqnarray}
Security is defined with the normalized mutual information between \textit{M} and \textit{Z}:
\begin{eqnarray}
\textbf{Weak Security}: \mathop {\lim }\limits_{k \to \infty } \frac{{I(M;Z)}}{k} = 0\
\end{eqnarray}
Secrecy must assures that \textit{Z} does not provide sufficient information about \textit{M}. Maurer in~\cite{mau94,mau20} proves the weakness of conventional notion of security (Eq.~2). Maurer defined strong security condition in~\cite{mau94}:
\begin{eqnarray}
\textbf{Strong Security}: \mathop {\lim }\limits_{k \to \infty } I(M;Z) = 0
\end{eqnarray}
Both weak and strong security conditions are information-theoretic and not computational for which there is no limitation for computational power of adversary. Hence, Computational complexity of algorithm does not affect the security of the system.
\subsection{Prior Works}
In 1975, Wyner  considered the special setting where both $W_m$ and $W_w$ are discrete memoryless channels (DMCs) and $W_w$ is degraded with respect to $W_m$ and characterized such a system by $C_s$ which is called the secrecy capacity: For $\forall \varepsilon  > 0$, there exists a coding scheme with information rate
$R > {C_s} - \varepsilon$ that simultaneously satisfies both (1) and (2). On the other hand, at rates higher than $C_s$~\cite{wyn75} both (1) and (2) cannot be satisfied.

Since 1975, Wyner’s works have been applied and generalized to a wide variety of other research efforts such as Gaussian channels~\cite{che78}, broadcast channels~\cite{csi78}, and channels by considering constraints on the computational power of the adversary~\cite{che09,oza84}. However, a large number of these works are based on non-constructive coding that only demonstrate the existence of a coding scheme that achieves secrecy capacity without proposing specific encoding-decoding schemes. The concept of mutual entropy has also been implemented in a large amount of research efforts such as \cite{iranmanesh2016robustness,iranmaneshmutual}.

To the best of our knowledge, research efforts that propose encoding-decoding algorithms exist only for two special cases. The first is having a noiseless main channel and the adversary channel is binary erasure channel~(BEC). LDPC codes for the BEC, are presented in~\cite{sur10} and~\cite{tha07} and achieve secrecy capacity. The other case is when the Eve has limited computational power. This case has been investigated by Ozarow and Wyner in~\cite{oza84}.

\subsection{Assumptions and Settings}
In this paper, a coding scheme that is proposed that achieves the secrecy capacity. The assumption is that $W_m$ and $W_w$ are binary-input non-symmetric DMCs and $W_w$ is degraded with respect to $W_m$. Conditions of (1) and (2) could be satisfied using an encoding/decoding algorithms by polar with the computational complexity of $O(NlogN)$. Our construction is inspired by the studies in polar codes recently investigated by Arkan~\cite{ari09}.

It is proved in~\cite{ari09} that polar codes can achieve the capacity of binary-input DMCs with low coding complexity. This proof is based on channel polarization. Arikan in~\cite{ari09} considered the channels observed by each of the individual bits during transformation. The channels seen by individual bits are called bit-channels. It is shown in~\cite{ari09} that by increasing the block length the bit-channels become either noiseless channels or the complete noisy ones. The bit-channels related to noiseless channels are called good channels and the other ones are called bad channels. The key result of~\cite{ari09} is that the portion of good channels becomes the capacity of \textit{W} while \textit{N} goes to infinity.

The main idea of our construction is as follows: The random bits will be transmitted over bit-channels which are good for both Bob and Eve. Message bits are transmitted over the bit-channels which are bad for Eve but good for Bob, and finally, a fixed vector is sent over bit-channels that are bad for all the parties. We prove that there exists a sequence of frozen vector that our coding scheme satisfies the reliability and security condition.
\subsection*{Organization}
In Section~\ref{section:Symmetric Concept and Secrecy Capacity}, we represent relevant concepts related to wiretap channels to provide a representation of the secrecy capacity in the setting where $W_m$ and $W_w$ are DMC and $W_w$ is degraded with respect to $W_m$. Also, the notion of symmetric channels and secrecy capacity are defined. Section~\ref{section:Polar Coding} is devoted to polar codes and theorems necessary for our proofs. We represent The proposed coding scheme in Section~\ref{section:Coding Scheme for achieving Secrecy Capacity} and we prove the security and reliability of proposed scheme. We also prove that the code rate approaches to secrecy capacity. In Section~\ref{section:Extending proofs to the whole capacity-equivocation region}, we prove that the equivocation rate for proposed coding scheme approaches to its upper bound in the whole capacity-equivocation region. In section~\ref{section:Simulation results} the simulation results are presented. Simulation results show that equivocation rate achieves its upper bound in the whole capacity-equivocation region.
\subsection*{Notations}
Random variables are denoted by upper case letters, the samples by the corresponding lower case letters. Calligraphic font represents the alphabet set of related random variable. $|{\cal X}|$ is the alphabet size. Notation ${A^N}$ is vector A of length N. $P_X$ is the distribution of X. If $f(x)$ and $g(x)$ are defined on a subset of real numbers then
$f(x) = O(g(x))$, if for large $x$ there exists a constant number M for which the inequality 
$f(x) \le M(g(x))$ holds. $a_1^N$ is vector $({a_1},{a_2},...,{a_N})$ and notation ${a_A}$ represents the sub vector $\left( {{a_i}:{\rm{ }}i \in A} \right)$.
$C_W$ represents the capacity of the channel W:
\begin{eqnarray}
C = \mathop {\max }\limits_{{P_X}} I(X;Y).
\end{eqnarray}
$I_W$ is the symmetric capacity of the channel W and for general channels, it is defined as:
\begin{eqnarray}
I(W) = \sum\limits_{y \in {\cal Y}} {\sum\limits_{x \in {\cal X}} {\frac{1}{{|{\cal X}|}}} } W(y|x)\log \frac{{W(y|x)}}{{{\textstyle{1 \over {|{\cal X}|}}}\sum\limits_{x' \in {\cal X}} {W(y|x')} }} 
\end{eqnarray}
This is the maximum achievable rate when all channel inputs $x$ are used with the same probability. If the maximizing distribution $P_X$ in (5) is the uniform distribution, then the symmetric capacity is equal to the capacity. $log(.)$ is based on 2 in the rest of the paper.
\subsection{Related works}
The works by Hof and Shamai~\cite{hof10} and Mahdavifar and Vardy~\cite{mah11} assume binary-input channels for the symmetric settings. In this work, we assume non-symmetric binary-input channels. Work in~\cite{mah11} considers only achieving secrecy capacity, but we prove that the proposed scheme for non-symmetric setting achieves all capacity-equivocation region. In~\cite{mah11} there is no assumption on the distribution of message for proving security condition which is a fair assumption on message M. But we consider uniform distribution since it is the necessary condition for approaching secrecy capacity. In~\cite{torfi17} the non-binary setting is investigated. However, no experimental results are presented. In this work, we also present simulation results of equivocation at Eve (using randomly chosen frozen vector) in BECs which measures the secrecy.
\section{Symmetric Concept and Secrecy Capacity}\label{section:Symmetric Concept and Secrecy Capacity}
In this section we review the works in~\cite{csi78} and~\cite{s77} to provide the notion of the secrecy capacity $C_s$. Our discussion is limited to the binary discrete memoryless channels with finite input-output alphabet size. Such channel is a triple $({\cal X},{\cal Y},W(y|x))$ in which ${\cal X}$ and ${\cal Y}$ are finite input-output alphabet of channel and \textit{W} is a $\left| {\cal X} \right| \times \left| {\cal Y} \right|$ matrix with $W(y|x)$ as entries. $W(y|x)$ is the probability of observing $y \in {\cal Y}$ if $x \in {\cal X}$ is sent.

A matrix \textit{W} is symmetric if the rows and columns of \textit{W} are permutations of each other, respectively. A channel $\left| {\cal X} \right| \times \left| {\cal Y} \right|$ is symmetric if \textit{W} is a symmetric matrix. Following~\cite{ari09} and~\cite{s77}, \textit{W} is weakly symmetric if the columns of \textit{W} can be split into parts such that each part forms a symmetric matrix.

For channel $W_b$ with input alphabet ${\cal Y}$ the wiretapper's channel is stochastically degraded with respect to the main channel, if it holds the following equation:
\begin{eqnarray}
{W_w}(z|x) = \sum\limits_{y' \in {\cal Y}} {{W_m}(y'|x)} {W_b}(z|y'){\rm{   }}\forall x,z 
\end{eqnarray}
If the channel transition probability factorizes as:
\begin{eqnarray}
W\left( {y,z|x} \right) = W\left( {y|x} \right){\rm{W}}\left( {z|y} \right),
\end{eqnarray}
the wiretapper's channel is called physically degraded with respect to the main channel. Since the capacity-equivocation region only depends on the marginal probabilities, the capacity-equivocation region for physically degraded and stochastically degraded wiretap channels is the~\cite{csi78} :
\begin{eqnarray}
\mathop U\limits_{{{\rm{P}}_{\rm{X}}}{{\rm{P}}_{{\rm{YZ}}\mid {\rm{X}}}}} \left\{ {\begin{array}{*{20}{c}}
{\begin{array}{*{20}{c}}
{\left( {R,{R_e}} \right):}\\
{0 \le {\rm{R}} \le {\rm{I}}\left( {{\rm{X}};{\rm{Y}}} \right)}
\end{array}}\\
{\begin{array}{*{20}{c}}
{0 \le {R_e} \le {\rm{R}}}\\
{{R_e} \le {\rm{I}}\left( {{\rm{X}};{\rm{Y}}} \right) - {\rm{I}}\left( {{\rm{X}};{\rm{Z}}} \right)}
\end{array}}
\end{array}} \right.
\end{eqnarray}
in which $R_e$ is equivocation rate and define by $\frac{1}{N}H(M|Z)$ when \textit{N} goes to infinity. In this case, the secrecy capacity is:
\begin{eqnarray}
{C_s} = \mathop {\max }\limits_{{P_X}} \{ I(X;Y) - I(X;Z)\}
\end{eqnarray}
If the wiretap channel is physically degraded to main channel then $X \to Y \to Z$ and $I(X;Z) \le I(X;Y)$. In this case if the same input distribution $P_X$ maximizes both $I(X;Z)$ and $I(X;Y)$, for instance when both $W_m$ and $W_w$ are symmetric channels, the capacity-equivocation region is given by:
\begin{eqnarray}
{R_e} \le R \le {C_{{W_m}}},0 \le {R_e} \le {C_{{W_m}}} - {C_{{W_w}}}
\end{eqnarray}
If \textit{W} is nonsymmetric, $C_{W_m}$ and $C_{W_w}$ are equal to $I_{W_m}$ and $I_{W_w}$. The secrecy capacity is:
\begin{eqnarray}
{C_s} = {C_{{W_m}}} - {C_{{W_w}}}
\end{eqnarray}
for the rest of the paper, degraded means stochastically degraded.
\section{Polar Coding}\label{section:Polar Coding}
In this section important notions of polar coding are defined which are used in our designs and proofs.
\subsection{Primitive Definitions}
Consider a binary-input discrete memoryless channel (B-DMC) given by $W(y|x)$ where $x \in {\cal X} = \{ 0,1\} $, $y \in {\cal Y}$ for finite set ${\cal Y}$. The \textit{N} uses of \textit{W} is denoted by ${W^N}(y_1^N|x_1^N)$. The symmetric capacity of a B-DMC is given by:
\begin{eqnarray}
I(W) = \sum\limits_{y \in {\cal Y}} {\sum\limits_{x \in {\cal X}} {\frac{1}{2}W(y|x)\log \frac{{2 \times W(y|x)}}{{W(y|0) + W(y|1)}}} } 
\end{eqnarray}
that is a special case of (5). The \textit{W} Bhattacharyya parameter is given by:
\begin{eqnarray}
Z\left( W \right) = \sum\limits_{y \in {\cal Y}} {\sqrt {W(y|0)W(y|1)} }
\end{eqnarray}
which measures the reliability of \textit{W} since it is an upper bound on the probability of \textit{ML} decision error on a single use of the channel.

Polar coding is introduced by Arikan~\cite{ari09}. The channel polarization phenomenon is used to construct codes that achieve the symmetric capacity $I(W)$ for B-DMC \textit{W} with encoding and decoding complexity that scales as $O(NlogN)$ with the block length . Channel polarization consists of two operations: channel combining and channel splitting. Let $u_1^N$ be the vector that supposed to be formed. The combined channel ${W_N}$ is represented by :
\begin{eqnarray}
{W_N}(y_1^N|u_1^N) = {W^N}(y_1^N|u_1^N{B_N}{F^{ \otimes N}})
\end{eqnarray}
Let  $F = \begin{bmatrix} 1& 0 \\1&1 \end{bmatrix}$, and let ${F^{ \otimes n}}$ denote the \textit{n-th} Kronecker power of \textit{F}. Let \textit{W} be arbitrary-input DMC, and the vector $U = ({U_1},{U_2},...,{U_N})$ be a block of $N = {2^n}$ bits chosen uniformly at random. Suppose U is encoded as $X = U{R_N}{F^{ \otimes n}}$, where ${R_N}$ is the bit-reversal permutation matrix. \textit{X} is sent through \textit{N} independent identical \textit{W}.

The channel splitting phase constructs \textit{N} binary-input channels from ${W_N}$, where the transformation is given by:
\begin{eqnarray}
W_N^{(i)}(y_1^N,u_1^{i - 1}|{u_i}) \equiv \sum\limits_{u_{i + 1}^N \in {{\cal X}^{N - i}}} {\frac{1}{{{2^{N - 1}}}}{W_N}(y_1^N|u_1^N)} 
\end{eqnarray}

Polar coding utilizes the polarization effect. It transmits data through the bitchannels for which $Z(W_N^{(i)})$ is close to 0. In~\cite{ari09} the polar code $(N,K,{\cal A},{u_{{{\cal A}^c}}})$ for B-DMC \textit{W} is defined by $x_1^N = u_1^N{B_N}{F^{ \otimes n}}$  where ${u_{{{\cal A}^c}}}$ is a given frozen vector and the information set ${\cal A}$ is chosen such that $|{\cal A}| = K$ and $Z(W_N^{(i)}) < Z(W_N^{(j)})$ for all $i \in {\cal A},j \in {{\cal A}^c}$. The frozen vector ${u_{{{\cal A}^c}}}$ is given to the decoder. Successive cancellation (\textit{SC}) decoder estimates the input as follows:for the frozen indices ${u_{{A^c}}} = {{\hat u}_{{A^c}}}$. For the remaining indices satisfying $i \in {\cal A}$;${{\hat u}_i} = 0$, if ${\mathop W\nolimits_N^{(i)} (y_1^N,\hat u_1^{i - 1}|0) \ge \mathop W\nolimits_N^{(i)} (y_1^N,\hat u_1^{i - 1}|1)}$ and ${{\hat u}_i} = 1$, otherwise.
\subsection{Polar coding ensemble}
According to polar coding, having a set of channels between encoder and decoder and sending information set on the good channels, is a coding scheme. Inputs to the other channels remain fix and are declared to the decoder. As the fraction of good channels is $I(W)$, the rate of $I(W)$ is achievable.

Mutual information $I({U_i};Y_1^N,U_1^{i - 1})$ corresponds to decoding $U_i$ with respect to the knowledge of $U_1^{i - 1}$ and output $Y_1^N$. In this case decoder should have the knowledge of $U_1^{i - 1}$ for decoding $U_i$. However, the decoder only knows $U_j$ where $j$ belongs to the indices of bad channels so for the other indices decoder should use the estimate ${\hat U_j}$, which could be incorrect. Successive cancellation decoder,decodes bits in clean order ${U_1},...,{U_N}$. For the moment polar code notation is represented which is used for defining polar coding ensemble and the rest of the paper.

\textbf{Definition 3.1} (\textit{Polar Coding~\cite{ari09}}):
Polar code ${{\cal P}}(N,{{\cal A}},{u_{{\cal F}}})$ for every ${{\cal A}} \subseteq \{ 1,...,N\} $ and ${u_{{\cal F}}} \in {{{\cal X}}^{|{{\cal F}}|}}$ is a linear code according to the following notation:
\begin{eqnarray}
{{\cal P}}(N,{{\cal A}},{u_{{\cal F}}}) = \{ x_1^N = u_1^N{G_N}:{u_{{{{\cal F}}^c}}} \in {{{\cal X}}^{|{{{\cal F}}^c}|}}\}
\end{eqnarray}
Code ${{\cal P}}(N,{{\cal A}},{u_{{\cal F}}})$ is constructed using a fixed vector ${u_{{\cal F}}}$ with indices set ${{\cal F}}$ and choosing from all possible vectors in indices ${{{\cal F}}^c}$ or ${{\cal A}}$. Notations ${{\cal P}}(N,{{\cal A}},{u_{{\cal F}}})$ and ${{\cal P}}(N,{{\cal A}},{u_{{{{\cal A}}^c}}})$ are equivalent.
${{\cal F}}$ is frozen set and its indices are called forzen indices. Also ${{\cal A}}$ is information set and its indices are called information indices. Using code ${{\cal P}}(N,{{\cal A}},{u_{{{{\cal A}}^c}}})$ is corresponds to sending $U_1^N$ on channel ${W_N}$ with a fixed ${u_{{\cal F}}}$ on the indices ${{\cal F}}$.

\textbf{Definition 3.2} (\textit{Polar Coding Ensemble~\cite{kor09}}): Polar code Ensemble ${{\cal P}}(N,{{\cal A}})$ for every ${{\cal A}} \subseteq \{ 1,...,N\} $ represents the Ensemble below:
\begin{eqnarray}
{{\cal P}}(N,{{\cal A}}) = \{ {{\cal P}}(N,{{\cal A}},{u_{{\cal F}}}):\forall {u_{{\cal F}}} \in {{{\cal X}}^{|{{\cal F}}|}}\} 
\end{eqnarray}
${P_{B,e}}({{\cal A}},{u_{{\cal F}}})$ represents the error probability of code block ${{\cal P}}(N,{{\cal A}},{u_{{\cal F}}})$ with uniform distribution assumption on all codewords. ${P_{B,e}}({{\cal A}})$ is the average block error probability of ensemble ${{\cal P}}(N,{{\cal A}})$ which is averaging ${P_{B,e}}({{\cal A}},{u_{{\cal F}}})$ on all possible choices of ${u_{{\cal F}}} \in {{{\cal X}}^{|{{\cal F}}|}}$ with equal probability.

\textbf{Lemma 3.1} (\textit{Average Block Error Probability Upper Bound}~\cite{kor09}): For a B-DMC W and information set ${{\cal A}}$, average block error probability over all possible choices of frozen bits can be bounded as follows:
\begin{eqnarray}
{P_{B,e}}({{\cal A}}) \le \sum\limits_{i \in {{\cal A}}} {Z(W_N^{(i)})}  
\end{eqnarray}
\subsection{Rate of polarization and achieving symmetric capacity}
\textbf{Theorem 3.1} (\textit{Rate of Convergence~\cite{ari09}}):for any B-DMC W for $N = {2^n}$ and $\delta  \in (0,1)$ :
\begin{eqnarray}
\mathop {\lim }\limits_{N \to \infty } \frac{{|i \in \{ 1,...,N\} :I(W_N^{(i)}) \in (1 - \delta ,1)|}}{N} = I(W) 
\end{eqnarray}
\begin{eqnarray}
\mathop {\lim }\limits_{N \to \infty } \frac{{|i \in \{ 1,...,N\} :I(W_N^{(i)}) \in (0,\delta )|}}{N} = 1 - I(W)
\end{eqnarray}
in order to derive the rate of channel polarization the random process ${{\cal Z}_n}$ defined in~\cite{ari09} and~\cite{tel09}:
 \begin{eqnarray}
\Pr \{ {{\cal Z}_n} \in (a,b)\} = \frac{{|i \in \{ 1,...,N\} :Z(W_{{2^n}}^{(i)}) \in (a,b)|}}{{{2^n}}}
\end{eqnarray}

\textbf{Theorem 3.2} (\textit{Polarization rate~\cite{tel09}, Theorem.1}):
For any B-DMC W and $0 < \beta  < 1/2$, $\mathop {\lim }\limits_{n \to \infty } \Pr \{ {{\cal Z}_n} < {2^{ - {2^n}^\beta }}\}  = I(W)$.

\textbf{Theorem 3.3} (\textit{\cite{tel09}, Theorem.2}): For any B-DMC W in which  $I(W) > 0$ and $ R < I(W)$, parameter $\beta  \in (0,1/2)$ is considered to be fixed. Block error probability of polar coding averaged over all possible choices of frozen bits satisfy the following equality:
\begin{eqnarray}
{P_{B,e}}({{\cal A}}) = O({2^{ - {N^\beta }}})
\end{eqnarray}
A lemma from~\cite{kor09} is used for realizing good channels and bad channels from each other.

\textbf{Lemma 3.2} (\textit{\cite{kor09}, Lemma 4.7}):
If $W:{\cal X} \to {\cal Y}$ and ${W_d}:{\cal X} \to {{\cal Y}_d}$ are two B-DMC W and $W_d$ is degraded with respect to $W$ then there exists a channel like ${W_b}:{\cal Y} \to {{\cal Y}_d}$ that ${W_d}({y_d}|x) = \sum\limits_{y \in {\cal Y}} {W(y|x){W_b}({y_d}|y)} $. In this condition ${W_d}_N^{(i)}$ is degraded with respect to ${W}_N^{(i)}$ and $Z({W_d}_N^{(i)}) \ge Z(W_N^{(i)})$.

This lemma implies that with assumption of degradation of wiretap channel with respect to main channel if a channel is good to Eve, it is good for Bob. Conversely if a channel is bad for Bob, it is bad for Eve too.
\subsection{Nested polar code}
We consider binary polar codes of block length $N = {2^n}$. Let ${\cal A}$ and ${\cal B}$ be two index sets such that ${\cal B} \subset {\cal A} \subset \{ 1,...,N\} $. Nested structure of polar codes comes from the cosets of a smaller subcodes. Consider the polar codes ${\cal P}(N,{\cal A},{u_{{{\cal A}^c}}})$ and ${\cal P}(N,{\cal B},[s,{u_{{{\cal A}^c}}}])$. Here $[s,{u_{{{\cal A}^c}}}]$ is a binary vector whose elements are equal to vector $s$ for the indices $i$ in ${\cal A}\backslash {\cal B}$, and otherwise they equal the corresponding elements in ${u_{{{\cal A}^c}}}$. ${{\cal A}^c}$ is a frozen set for both codes, but ${{\cal B}^c}$ is frozen only for ${\cal P}(N,{\cal B},[s,{u_{{{\cal A}^c}}}])$.

\textbf{Definition 3.3} (\textit{Nested Polar Code~\cite{and10}}):
Let ${G_N}$ be the generator matrix of polar code and let ${G_N}(I)$ be the submatrix composed of the rows of ${G_N}$ whose indices belong to the set $I$. The nested polar code ${\cal P}(N,{\cal A},{\cal B},{u_{{{\cal A}^c}}})$ is the set of codewords ${x^N}$ of the following form:
\begin{eqnarray}
{x^N} = {u_{\cal B}}{G_N}({\cal B}) \oplus {u_{{\cal A}\backslash {\cal B}}}{G_N}({\cal A}\backslash {\cal B}) \oplus {u_{{{\cal A}^c}}}{G_N}({{\cal A}^c})
\end{eqnarray}
The rate of the sub-codes ${\cal P}(N,{\cal B},[{u_{{\cal A}\backslash {\cal B}}},{u_{{{\cal A}^c}}}])$ equal $\frac{{|{\cal B}|}}{N}$, and the rate of the overall code ${\cal P}(N,{\cal A},{u_{{{\cal A}^c}}})$ equals $\frac{{|{\cal A}|}}{N}$. 
\section{Coding Scheme for achieving Secrecy Capacity}\label{section:Coding Scheme for achieving Secrecy Capacity}
In this section we represent a coding scheme and prove its security and reliability . Also we show that it achieves the rate of secrecy capacity.
\subsection{Secret Codebook}
A discrete memoryless wiretap channel is denoted by following notation:
\begin{eqnarray}
({\cal X},W(y,z|x),{\cal Y} \times {\cal Z})
\end{eqnarray}
finite sets ${\cal X}$,${\cal Y}$ and ${\cal Z}$ are input alphabet, main channel and wiretap channel alphabet in the corresponding order. The channel is assumed to be memoryless and time-invariant:
\begin{eqnarray}
W({y_i},{z_i}|x_1^i,y_1^{i - 1},z_1^{i - 1}) = W({y_i},{z_i}|{x_i})
\end{eqnarray}
Assume that the transmitter has a confidential message M which is to be transmitted to the receiver and to be hidden from the Eve. The secret codebook is defined as below:
\begin{enumerate}
\item secret message \textbf{M}. The transmitted messages are assumed to be uniformly distributed over message set ${\cal M}$. 
\item  encoding function $encod(.)$ at the transmitter which maps the secret messages to the transmitted symbols: for each $m \in {\cal M}$ $encod:m \to x_1^N$
\item  Decoding function $decod(.)$ at receiver which maps the received symbols to estimate of the message: $decod(y_1^N)={\hat m}$
\end{enumerate}

Reliability is measured by block error probability:
\begin{eqnarray}
{P_e} = \frac{1}{{|{\cal M}|}}\sum\limits_{m \in {\cal M}} {\Pr \{ decod(y_1^N) \ne m| \textit{m is sent})} 
\end{eqnarray}
Security is measured by the mutual information leakage rate to the eavesdropper:
\begin{eqnarray}
\frac{1}{N}I(M;Z_1^N)
\end{eqnarray}
The rate R is called achievable secrecy rate, if for any given $\varepsilon  > 0$, there exists a secret codebook such that:
\begin{eqnarray}
Rate  :\frac{1}{N}\log (|{\cal M}|) = R  
\end{eqnarray}
\begin{eqnarray}
Reliability :{P_{B,e}}({{{\cal A}}_m}) \le \varepsilon 
\end{eqnarray}
\begin{eqnarray}
Security :\frac{1}{N}\log (M;Z_1^N) \le \varepsilon 
\end{eqnarray}
Wiretap channel is degraded with respect to the main channel. Relation between the input and output of the main and wiretap channels depicted as following Markov chain: $U \to X \to (Y,Z)$

For sufficiently large N and $0 < \beta  < 1/2$ following sets are defined:
\begin{eqnarray}
{{{\cal A}}_m} = \{ i \in \{ 1,...,N\} :Z({W_m}_N^{(i)}) \le \frac{1}{N}{2^{ - {N^\beta }}}\} 
\end{eqnarray}
\begin{eqnarray}
{{{\cal A}}_w} = \{ i \in \{ 1,...,N\} :Z({W_w}_N^{(i)}) \le \frac{1}{N}{2^{ - {N^\beta }}}\}  
\end{eqnarray}
According to the defined sets, ${{{\cal A}}_m}$ is the good channel indices for the main channel and ${{{\cal A}}_w}$ corresponds to the good channel indices for the wiretap channel. Considering the polar coding definition and lemma 3.2 it is concluded that ${{\cal F}_m} \subseteq {{\cal F}_w},{{\cal A}_w} \subseteq {{\cal A}_m}$. We consider nested polar code ${\cal P}(N,{{\cal A}_m},{{\cal A}_w},{u_{{{\cal F}_m}}})$ defined in Section~IV.D. The main objective is to form $u_1^N$ based on defined indices sets.
$u_1^N$ is the vector that multiplied by generator matrix and forms transmitted codeword. Figure.1 demonstrates the relation between defined indices.
\begin{figure}[t]
\begin{center}
\includegraphics[scale=0.2]{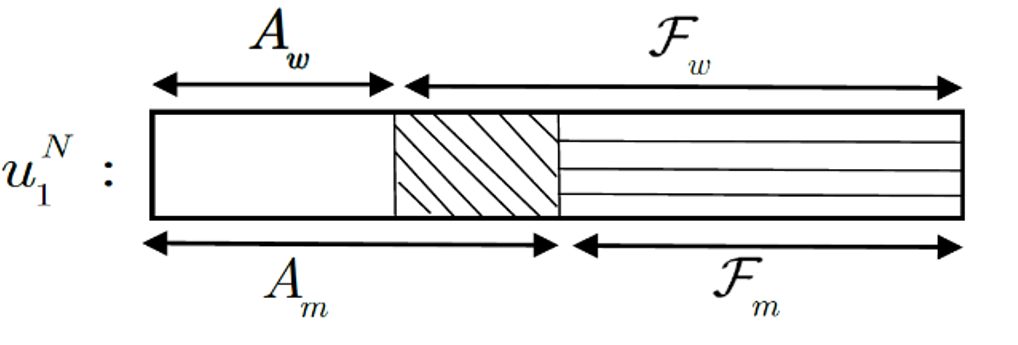}
\caption{Main channel and wiretap channel indices.}
\end{center}
\end{figure}
\subsection{Encoding Algorithm}
Secret message is mapped on the vector $V_m$ and random vector $V_r$ is generated with uniformly random distribution. The vector $u_1^N$ is formed as below:
\begin{enumerate}
\item Information bits are sent over the indices which are good for the main channel and bad for the wiretap channel. This concept could be shown by ${u_{{{\cal A}_m}\backslash {{\cal A}_w}}} = {u_{{{\cal F}_w}\backslash {{\cal F}_m}}} = {V_m}$ message length is $k$ and $|{V_m}| = |{{\cal A}_m}| - |{{\cal A}_w}| = k$. Message distribution is uniform and therefore $\log |{\cal M}| = k$.
\item We send random bits over indices that belong to good channel for both main and wiretap channels as ${u_{{{\cal A}_w}}} = {V_r}$.
\item Over bad channels for both (${u_{{{\cal F}_m}}}$), we send a frozen vector assumed to be chosen from all possible choices ${u_{{{\cal F}_m}}} \in {{{\cal X}}^{|{{{\cal F}}_m}|}}$ and given to decoder of Bob and Eve. Coding scheme is built over $\forall {u_{{{\cal F}_m}}} \in {{{\cal X}}^{|{{{\cal F}}_m}|}}$ and we should prove that there exists a specific frozen vector like ${u_{{{\cal F}_m}}}$ for which our coding scheme satisfies the reliability and security condition and additionally achieves the secrecy capacity.
\end{enumerate}
\subsection{Decoding}
Decoding should satisfy reliability and coding rate.
\subsubsection{Reliability}
$V_m$ and $V_r$ are defined over good indices of main channel, thus according to theorem 3.3, both could be decoded using SC decoding with probability of error ${P_{B,e}}({{{\cal A}}_m}) = O({2^{ - {N^\beta }}})$ (averaging over all possible choices of ${u_{{{\cal F}_m}}}$). Therefore, the reliability condition is satisfied.
\subsubsection{Rate}
Message distribution is uniformly at random. That leads to $\log |{\cal M}| = |{V_m}| = k$, consequently for sufficiently large \textit{N}:
\begin{eqnarray}
R = \frac{{|{V_m}|}}{N} = \frac{{|{{\cal A}_m}| - |{{\cal A}_w}|}}{N} = I({W_m}) - I({W_w}) = {C_s}
\end{eqnarray}
Thus, the coding scheme achieves the secrecy capacity.
\subsection{Security Proof}
Since the scheme is formed over all possible choices of frozen bits (polar coding ensemble), mutual information between message and Eve evaluated using one random chosen vector ${u_{{{\cal F}_m}}}$ over the whole set ${u_{{{\cal F}_m}}} \in {{{\cal X}}^{|{{{\cal F}}_m}|}}$. After choosing ${u_{{{\cal F}_m}}}$, we fix it and ultimately we prove that there exists such ${u_{{{\cal F}_m}}}$. The decoding error probability of Eve has been evaluated over the ensemble in average sense.
\begin{eqnarray}
I(M;Z_1^N|{U_{{{\cal F}_m}}}) = I({V_m};Z_1^N|{U_{{{\cal F}_m}}})
\end{eqnarray}
\begin{eqnarray}
 = I({V_m},{V_r};Z_1^N|{U_{{{\cal F}_m}}}) - I({V_r};Z_1^N|{V_m},{U_{{{\cal F}_m}}})
\end{eqnarray}
\begin{eqnarray}
 = I(U_1^N;Z_1^N) - I({V_r};Z_1^N|{V_m},{U_{{{\cal F}_m}}})
\end{eqnarray}
\begin{eqnarray}
 = I(U_1^N;Z_1^N) - H({V_r}) + H({V_r}|{V_m},{U_{{{\cal F}_m}}},Z_1^N)
\end{eqnarray}
\begin{eqnarray}
 \le I(X_1^N;Z_1^N) - H({V_r}) + H({V_r}|{V_m},{U_{{{\cal F}_m}}},Z_1^N)
\end{eqnarray}
\begin{eqnarray}
 \le NI({W_w}) - |{{\cal A}_w}| + H({V_r}|{V_m},{U_{{{\cal F}_m}}},Z_1^N)
\end{eqnarray}
Equation (35) is concluded from the chain rule of mutual information and (36) is a consequence of the following:
\begin{eqnarray}
\begin{array}{l}
I(U_1^N;Z_1^N) = I({U_{{V_m} \cup {V_r} \cup {U_{{{\cal F}_m}}}}};Z_1^N)\\
= I({V_m},{V_r},{U_{{{\cal F}_m}}};Z_1^N)\\
= I({U_{{{\cal F}_m}}};Z_1^N) + I({V_m},{V_r};Z_1^N|{U_{{{\cal F}_m}}})\\
= I({V_m},{V_r};Z_1^N|{U_{{{\cal F}_m}}})
\end{array}
\end{eqnarray}
The last equality in (40) is derived from $I({U_{{{\cal F}_m}}};Z_1^N)$ being equal to zero, since ${u_{{{\cal F}_m}}}$ is sent over bad channels for both main and wiretap channels. Equation (37) follows form the independence of ${V_r},{V_m}$, and ${U_{{{\cal F}_m}}}$. (38) comes from data processing inequality: $M \to U \to X \to (Y,Z)$. Below inequalities lead to (39):
\begin{eqnarray}
\begin{array}{l}
I(X_1^N;Z_1^N) = H(Z_1^N) - H(Z_1^N|X_1^N)\\
 = H(Z_1^N) - \sum\limits_{i = 1}^N {H({Z_i}|{X_i})} \\
 \le \sum {(H({Z_i})}  - H({Z_i}|{X_i})) = \sum\limits_{i = 1}^N {I({X_i};{Z_i})}  \le NI({W_w})
\end{array} 
\end{eqnarray}
According to (39) to find an upper bound for $I(M;Z_1^N|{U_{{{\cal F}_m}}})$, an upper bound for $H({V_r}|{V_m},{U_{{{\cal F}_m}}},Z_1^N)$ should be found.

\textbf{Lemma 4.1}: For a nonsymmetric binary-input channel the ensemble of polar code was defined which means defining polar code over all possible choices of frozen bits. For proposed coding scheme defined over the ensemble there exists a frozen vector ${u_{{{\cal F}_m}}}$, which satisfies the following inequality:
\begin{eqnarray}
 H({V_r}|{V_m},{U_{{{\cal F}_m}}},Z_1^N)/N \le \varepsilon 
\end{eqnarray}
\begin{proof}
we define the following error event:
\begin{eqnarray}
E = \left\{ {\begin{array}{*{20}{c}}
{\textbf{1          } {{\hat V}_r} \ne {V_r}}\\
{\textbf{0          } {{\hat V}_r} = {V_r}}
\end{array}} \right.
\end{eqnarray}
Random vector is sent over good channels for both main channel and wiretap channel. Therefore, ${P_e} = {P_{B,e}}({{\cal A}_w})$ and:
\begin{eqnarray}
\begin{array}{l}
{P_e} = P(E = 1) = \Pr ({{\hat V}_r} \ne {V_r})\\
\le \sum\limits_{i \in {{\cal A}_w}} {Z({W_w}_N^{(i)})}  \le {2^{ - {N^\beta }}}
\end{array}
\end{eqnarray}
Since coding scheme is defined over all possible choices of ${u_{{{\cal F}_m}}} \in {{{\cal X}}^{|{{{\cal F}}_m}|}}$ and the error probability in average sense is smaller than its upper bound, there exists a specific frozen vector  ${u_{{\cal F}_m}}$ which is in set ${{{\cal X}}^{|{{{\cal F}}_m}|}}$. By choosing it, the error probability does not exceed the upper bound ${2^{ - {N^\beta }}}$. The term $H(E,{V_r}|{V_m},{U_{{{\cal F}_m}}},Z_1^N)$ is expanded in two ways:
\begin{eqnarray}
\begin{array}{l}
H(E,{V_r}|{V_m},{U_{{{\cal F}_m}}},Z_1^N) = \\
\\
H({V_r}|{V_m},{U_{{{\cal F}_m}}},Z_1^N) + H(E|{V_r},{V_m},{U_{{{\cal F}_m}}},Z_1^N) = \\
\\
H({V_r}|E,{V_m},{U_{{{\cal F}_m}}},Z_1^N) + H(E|{V_m},{U_{{{\cal F}_m}}},Z_1^N)
\end{array}
\end{eqnarray}
$H(E|{V_r},{V_m},{U_{{{\cal F}_m}}},Z_1^N)$ equals zero, and consequently:
\begin{eqnarray}
\begin{array}{l}
H({V_r}|{V_m},{U_{{{\cal F}_m}}},Z_1^N) = \\
H({V_r}|E,{V_m},{U_{{{\cal F}_m}}},Z_1^N) + H(E|{V_m},{U_{{{\cal F}_m}}},Z_1^N)
\end{array}
\end{eqnarray}
\begin{eqnarray}
\begin{array}{l}
\le H({V_r}|E,{V_m},{U_{{{\cal F}_m}}},Z_1^N) + H(E)
\end{array}
\end{eqnarray}
To find an upper bound for $H({V_r}|E,{V_m},{U_{{{\cal F}_m}}},Z_1^N)$:
\begin{eqnarray}
\begin{array}{l}
H({V_r}|E,{V_m},{U_{{{\cal F}_m}}},Z_1^N) = \\
\sum\limits_{i = 0}^1 {P(E = i)H({V_r}|E = i,{V_m},{U_{{{\cal F}_m}}},Z_1^N)} \\
= P(E = 1)H({V_r}|E = 1,{V_m},{U_{{{\cal F}_m}}},Z_1^N)\\
+ (1 - P(E = 1)) \times 0\\
\to H({V_r}|E,{V_m},{U_{{{\cal F}_m}}},Z_1^N)\\
= P(E = 1)H({V_r}|E = 1,{V_m},{U_{{{\cal F}_m}}},Z_1^N)\\
\le P(E = 1)H({V_r}) = {P_e}|{{\cal A}_w}|
\end{array}
\end{eqnarray}
Considering (47) and (48):
\begin{eqnarray}
 H({V_r}|{V_m},{U_{{{\cal F}_m}}},Z_1^N) \le H(E) + {P_e}|{{\cal A}_w}|
\end{eqnarray}
\begin{eqnarray}
\to H({V_r}|{V_m},{U_{{{\cal F}_m}}},Z_1^N)/N \le \frac{1}{N}(H(E) + {P_e}|{{\cal A}_w}|)
\end{eqnarray}
\begin{eqnarray}
\begin{array}{l}
\to H({V_r}|{V_m},{U_{{{\cal F}_m}}},Z_1^N)/N\\
\le \frac{1}{N}(H({2^{ - {N^\beta }}}) + {2^{ - {N^\beta }}}|{{\cal A}_w}|)
\end{array} 
\end{eqnarray}
\begin{eqnarray}
\begin{array}{l}
\to \mathop {\lim }\limits_{N \to \infty } H({V_r}|{V_m},{U_{{{\cal F}_m}}},Z_1^N)/N\\
\le \mathop {\lim }\limits_{N \to \infty } \frac{1}{N}(H({2^{ - {N^\beta }}}) + {2^{ - {N^\beta }}}|{{\cal A}_w}|)
\end{array}
\end{eqnarray}
\begin{eqnarray}
= \mathop {\lim }\limits_{N \to \infty } {2^{ - {N^\beta }}}I({W_w}) = 0
\end{eqnarray}
\end{proof}
Considering lemma 4.1 and (39):
\begin{eqnarray}
\frac{1}{N}I(M;Z_1^N|{U_{{{\cal F}_m}}}) \le I({W_w}) - |{{\cal A}_w}|/N\mathop  \approx \limits^{N \to \infty } 0
\end{eqnarray}
Therefore, the security of coding scheme is proved, since $\mathop {\lim }\limits_{N \to \infty } I(M;Z)/N$ is equivalent to $\mathop {\lim }\limits_{k \to \infty } I(M;Z)/k$. Equation (54) holds since $I({W_w})\mathop  \equiv \limits^{N \to \infty } |{{\cal A}_w}|/N$.
\section{Extending proofs to the whole capacity-equivocation region}\label{section:Extending proofs to the whole capacity-equivocation region}
In this section we show that for the proposed scheme the equivocation rate approaches its upper bound for rates higher than secrecy capacity $C_s$. According to (10), for rates higher that $C_s$, and for binary-input nonsymmetric channels the upper bound for equivocation rate equals $C_s=I_{W_m}-I_{W_w}$ . We expand $I(X_1^N,M;Z_1^N)$ in two forms:
\begin{eqnarray}
\begin{array}{l}
I(X_1^N,M;Z_1^N|{U_{{{\cal F}_m}}}) = 
I(X_1^N;Z_1^N|{U_{{{\cal F}_m}}}) \\ + I(M;Z_1^N|X_1^N,{U_{{{\cal F}_m}}})
\end{array}
\end{eqnarray}
\begin{eqnarray}
 = I(M;Z_1^N|{U_{{{\cal F}_m}}}) + I(X_1^N;Z_1^N|M,{U_{{{\cal F}_m}}})
\end{eqnarray}
In (55) $I(M;Z_1^N|X_1^N,{U_{{{\cal F}_m}}})$ equals to zero ($M \to X \to Y \to Z$). From (55) and (56):
\begin{eqnarray}
\begin{array}{l}
I(M;Z_1^N|{U_{{{\cal F}_m}}}) = I(X_1^N;Z_1^N|{U_{{{\cal F}_m}}})\\
 - I(X_1^N;Z_1^N|M,{U_{{{\cal F}_m}}})
\end{array}
\end{eqnarray}
Equivocation rate $H(M|Z_1^N,{U_{{{\cal F}_m}}})/N$ is expanded as following equations:
\begin{eqnarray}
\frac{{H(M|Z_1^N,{U_{{{\cal F}_m}}})}}{N} = \frac{{H(M|{U_{{{\cal F}_m}}}) - I(M;Z_1^N|{U_{{{\cal F}_m}}})}}{N}=
\end{eqnarray}
\begin{eqnarray}
\frac{{H(M|{U_{{{\cal F}_m}}}) + I(X_1^N;Z_1^N|M,{U_{{{\cal F}_m}}}) - I(X_1^N;Z_1^N|{U_{{{\cal F}_m}}})}}{N}
\end{eqnarray}
(59) concluded using (58) and (57). From (59):
\begin{eqnarray}
\begin{array}{l}
H(M|{U_{{{\cal F}_m}}})/N + H(X_1^N|M,{U_{{{\cal F}_m}}})/N\\
 - H(X_1^N|Z_1^N,M,{U_{{{\cal F}_m}}})/N - I(X_1^N;Z_1^N|{U_{{{\cal F}_m}}})/N
\end{array} 
\end{eqnarray}
\begin{eqnarray}
\begin{array}{l}
H(M,X_1^N|{U_{{{\cal F}_m}}})/N - H(X_1^N|Z_1^N,M,{U_{{{\cal F}_m}}})/N\\
 - I(X_1^N;Z_1^N|{U_{{{\cal F}_m}}})/N
\end{array}
\end{eqnarray}
\begin{eqnarray}
\begin{array}{l}
H(X_1^N|{U_{{{\cal F}_m}}})/N - H(X_1^N|Z_1^N,M,{U_{{{\cal F}_m}}})/N\\
 - I(X_1^N;Z_1^N|{U_{{{\cal F}_m}}})/N
\end{array}
\end{eqnarray}
\begin{eqnarray}
 \ge \frac{{|{{\cal A}_m}|}}{N} - \frac{{H(X_1^N|Z_1^N,M,{U_{{{\cal F}_m}}})}}{N} - {I_{{W_w}}}
\end{eqnarray}
Equation (62) is derived from the Markov chain $M \to U \to X \to (Y,Z)$. (63) follows from $I(X_1^N;Z_1^N|{U_{{{\cal F}_m}}})/N \le {I_{{W_w}}}$ and $H(X_1^N|{U_{{{\cal F}_m}}})/N = |{{\cal A}_m}|/N$.

Inequality $I(X_1^N;Z_1^N|{U_{{{\cal F}_m}}})/N \le {I_{{W_w}}}$ holds:
\begin{eqnarray}
\begin{array}{l}
I(X_1^N;Z_1^N|{U_{{{\cal F}_m}}}) = H(Z_1^N|{U_{{{\cal F}_m}}}) - H(Z_1^N|X_1^N,{U_{{{\cal F}_m}}})\\
 = H(Z_1^N|{U_{{{\cal F}_m}}}) - \sum\limits_{i = 1}^N {H({Z_i}|{X_i},{U_{{{\cal F}_m}}})} \\
 \le \sum {(H({Z_i}|{U_{{{\cal F}_m}}})}  - H({Z_i}|{X_i},{U_{{{\cal F}_m}}}))\\
 = \sum\limits_{i = 1}^N {I({X_i};{Z_i}|{U_{{{\cal F}_m}}})}  \le NI({W_w})
\end{array}
\end{eqnarray}
$H(X_1^N|{U_{{{\cal F}_m}}})/N = |{{\cal A}_m}|/N$ holds since transmitted codewords are uniformly distributed and for specific ${u_{{{\cal F}_m}}}$ the entropy of sequence $X_1^N$ equals to the number of good indices for the main channel.
According to analyzes:
\begin{eqnarray}
\begin{array}{l}
H(M|Z_1^N,{U_{{{\cal F}_m}}})/N \ge |{{\cal A}_m}|/N\\
 - H(X_1^N|Z_1^N,M,{U_{{{\cal F}_m}}})/N - {I_{{W_w}}}
\end{array}
\end{eqnarray}
Equation $H(X_1^N|Z_1^N,M,{U_{{{\cal F}_m}}}) = H({V_r}|Z_1^N,M,{U_{{{\cal F}_m}}})$ and lemma 4.1 prove that there exists a sequence of frozen bits for which the following inequality holds:
\begin{eqnarray}
H(X_1^N|Z_1^N,M,{U_{{{\cal F}_m}}}) \le N\varepsilon 
\end{eqnarray}
Therefore,
\begin{eqnarray}
H(M|Z_1^N,{U_{{{\cal F}_m}}})/N \ge |{{\cal A}_m}|/N - \varepsilon  - {I_{{W_w}}}
\end{eqnarray}
\begin{eqnarray}
 \to \frac{{H(M|Z_1^N)}}{N} \ge \frac{{H(M|Z_1^N,{U_{{{\cal F}_m}}})}}{N} \ge \frac{{|{{\cal A}_m}|}}{N} - \varepsilon  - {I_{{W_w}}}
\end{eqnarray}
\begin{eqnarray}
 \to \frac{{H(M|Z_1^N)}}{N} \ge \frac{{|{{\cal A}_m}|}}{N} - \varepsilon  - {I_{{W_w}}}
\end{eqnarray}
And for sufficiently large N:
\begin{eqnarray}
\mathop {Lim}\limits_{N \to \infty } \frac{{H(M|Z_1^N)}}{N} = {I_{{W_m}}} - {I_{{W_w}}}
\end{eqnarray}
\section{Simulation results}\label{section:Simulation results}
In this section results for calculation of equivocation at Eve are presented to support the theoretic proofs. We show that the equivocation rate achieves its upper bound for all rates. First, equivocation at Eve $H(M|{Z^N})$ has been introduced. Then, results has been presented. In all the settings, message has uniform distribution and both main and wiretap channels are BEC. BEC channels are of great interest since there are recursive equations to calculate the Bhattacharrya parameter~\cite{ari09}.
\subsection{Equivocation at Eve}
The measure for security is $I(M;{Z^N})/N$. It is expanded to:
\begin{eqnarray}
\frac{{I(M;{Z^N})}}{N} = \frac{{H(M)}}{N} - \frac{{H(M|{Z^N})}}{N}
\end{eqnarray}
\begin{eqnarray}
 = \frac{k}{N} - \frac{{H(M|{Z^N})}}{N} = R - \frac{{H(M|{Z^N})}}{N}
\end{eqnarray}
To derive $I(M;{Z^N})/N$, calculating $H(M|{Z^N})$ is sufficient. We propose the following lemma which is an extension to~\cite{and10} Lemma 4.1.

\textbf{lemma 6.1}: Assume that the nested polar code is considered ${\cal P}(N,{\cal A},{\cal B},{u_{\cal F}})$ (${u_{\cal F}}$ is a randomly chosen vector and fixed after selecting), and ${\cal B} \subset {\cal A}$. ${H_T}$ is the parity check matrix for overall code (${\cal P}(N,{\cal A})$ in polar coding) and ${H_S}$ is the parity check matrix for sub-code ${\cal P}(N,{\cal B})$, and channel is BEC. Then, equivocation at Eve is calculated as following:
\begin{eqnarray}
H(M|{Z^N}) = Rank({\mathord{\buildrel{\lower3pt\hbox{$\scriptscriptstyle\frown$}} 
\over H} _S}(\varepsilon )) - Rank({\mathord{\buildrel{\lower3pt\hbox{$\scriptscriptstyle\frown$}} 
\over H} _T}(\varepsilon ))
\end{eqnarray}
$\mathord{\buildrel{\lower3pt\hbox{$\scriptscriptstyle\frown$}} 
\over H} (\varepsilon )$ is the matrix formed by columns of $H$ that belong to erased positions.
\begin{proof}
\begin{eqnarray}
\begin{array}{l}
I(M;{X^N}|{Z^N}) = H(M|{Z^N}) - H(M|{X^N},{Z^N})\\
 = H({X^N}|{Z^N}) - H({X^N}|M,{Z^N})\\
 \to H(M|{Z^N}) = H({X^N}|{Z^N}) - H({X^N}|M,{Z^N})
\end{array}
\end{eqnarray}
$H(M|{X^N},{Z^N})$ is equal to zero since knowing transmitted codeword results in the message to be realized. Channels are BEC. Therefore, for the received \textit{Z}, \textit{X} is explicit with some erased symbols. Transmitted vector is  built as ${x^N} = {u_{\cal A}}{G_N}({\cal A}) \oplus {u_{\cal F}}{G_N}({\cal F})$. ${G_T} = {G_N}({\cal A} \cup {{\cal F}_{non - zero}})$ is the generator matrix of polar code formed from the rows of the mother generator matrix $G_N$ that belongs to information indices and nonzero positions of frozen vector.
$H({X^N}|{Z^N})$ corresponds to overall code ${\cal P}(N,{\cal A})$, consequently for a received \textit{Z} parity check equation hold: ${\mathord{\buildrel{\lower3pt\hbox{$\scriptscriptstyle\frown$}} 
\over H} _T}.(x_\varepsilon ^T) = 0$. Therefore:
\begin{eqnarray}
{\mathord{\buildrel{\lower3pt\hbox{$\scriptscriptstyle\frown$}} 
\over H} _T}(\varepsilon )x_\varepsilon ^T + {\mathord{\buildrel{\lower3pt\hbox{$\scriptscriptstyle\frown$}} 
\over H} _T}({\varepsilon ^c})x_{{\varepsilon ^c}}^T = 0
\end{eqnarray}
Equation (75) holds for unknown $x_\varepsilon ^T$.It has ${2^{|\varepsilon|-Rank({{\mathord{\buildrel{\lower3pt\hbox{$\scriptscriptstyle\frown$}} 
\over H}}_T}(\varepsilon ))}}$ solutions with equal probabilities, since codewords are equal likely. $|\varepsilon |$ is the number of erased position. Consequently $H(X|{Z^N}) = |\varepsilon | - Rank({\mathord{\buildrel{\lower3pt\hbox{$\scriptscriptstyle\frown$}} 
\over H} _T}(\varepsilon ))$.

It can also be proven that $H(X|M,{Z^N}) = |\varepsilon|-Rank({\mathord{\buildrel{\lower3pt\hbox{$\scriptscriptstyle\frown$}}\over H}_S}(\varepsilon))$. Therefore, it concludes (73).
\end{proof}
\subsection{Results and parity check matrix calculation}
Table \ref{table_er} presents the results of calculating equivocation rate and normalized mutual information for ${\varepsilon _M} = 0.25$ and ${\varepsilon _W} = 0.5$, with respect to changing the rate.

The parity check matrix of any of overall codes or sub-codes could not be calculated from the generator matrix directly, since the generator matrix does not have a standard form. For instance for overall code ${\cal P}(N,{\cal A})$ after elementary row operation the generator matrix could be turn to reduced row echelon form and standard form of generator matrix is derived as ${G_T} = {G_N}({\cal A} \cup {{\cal F}_{non - zero}}) = [{I_{|{\cal A} \cup {{\cal F}_{non - zero}}|}}|{P_1}]$.
Then, the parity check matrix could be calculated as ${H_T} = [{P_1}^T|{I_{N - |{\cal A} \cup {{\cal F}_{non - zero}}|}}]$.


\begin{table}[h]
\caption[Table caption text]{Equivocation Rate}
\label{table_er}
\begin{center}
\addtolength{\tabcolsep}{1pt}
\begin{tabular}{ccc}
\toprule 
Rate & ${R_e} = H(M|{Z^N})/N$ &$I(M;{Z^N})/N$\\ 
\hline
\midrule

0.05 & 0.0488 & 0.0012\\

\rowcolor{black!10}0.1 & 0.0988 & 0.0053\\

0.15 & 0.1475 & 0.0025\\

\rowcolor{black!10}0.2 & 0.1992 & 0.0008\\

0.25 & 0.2425 & 0.0075\\

\rowcolor{black!10}0.3 & 0.2480 & 0.052\\

0.4 & 0.2485 & 0.1515\\

\rowcolor{black!10}0.5 & 0.2490 & 0.251\\

0.6 & 0.2492 & 0.3508\\
\bottomrule
\end{tabular}
\end{center}

\end{table}

\section{Conclusion and discussion}\label{section:Conclusion and discussion}
In this paper we considered binary-input non-symmetric wiretap channels. We proved that there exists a frozen vector for which coding scheme satisfies reliability and security conditions and also code rate achieves secrecy capacity. We proved that the equivocation rate achieves its upper bound for all rates in non-symmetric channels. Our results extend to discrete memoryless channels with non-binary input. It is proved in~\cite{sas09} that channels with an input alphabet of prime size are polarized by the same transformation. If the alphabet is not of prime size, then splitting the input alphabet into prime subsets can solve the problem.

All the constructions in this paper are as explicit as the polar codes since only existence of a suitable frozen vector is proved and the method to choose it is not explored. Maurer-Wolf~\cite{mau20} proved that coding schemes that satisfy the weak security condition, can be also be converted to schemes that satisfy the strong security condition too. This is accomplished using information reconciliation and privacy amplification protocol~\cite{ben95}. Therefore, the proposed scheme could be extended to strong security using privacy amplification protocol.

Another possible further problem to explore is to construct codes when the wiretap channel is not degraded since degradation of wiretap channel is a sufficient but not necessary condition. Also, the proposed algorithm benefits from successive cancellation decoding, which depends on the past estimates. Therefore, if the estimates are incorrect, the error will propagate. A decoder can be implemented to overcome this issue. In addition, it worth investigating the possibility of the other decoding methods, such as belief propagation~\cite{hus09} and recursive-list decoding~\cite{dum06} being able to eliminate this deficiency. 

Recently,~\cite{hon13} has proposed an algorithm to overcome the polarization restrictions. The method propose the definition of pseudo-random frozen bits. This method can be utilized to generalize the proposed scheme for arbitrary discrete memoryless channels.


{\bibliographystyle{IEEEtran}
\bibliography{bib}
}

\end{document}